%% file: main.tex
\documentclass[reprint,floatfix]{revtex4-2}
\usepackage[utf8]{inputenc}
\usepackage[english]{babel}
\usepackage{amsmath}
\usepackage{amsfonts}
\usepackage{amssymb}
\usepackage{graphicx}
\usepackage{braket}
\usepackage{hyperref}
\usepackage{natbib}
\begin{document}
\preprint{APS/123-QED}

\title{Long Distance Nonlocality Test with Entanglement Swapping and Displacement-Based Measurements}

\author{Anders J. E. Bjerrum}
    \affiliation{Center for Macroscopic Quantum States (bigQ), Department of Physics, Technical University of Denmark, 2800 Kongens Lyngby, Denmark}
\author{Jonatan B. Brask}%
    \affiliation{Center for Macroscopic Quantum States (bigQ), Department of Physics, Technical University of Denmark, 2800 Kongens Lyngby, Denmark}
\author{Jonas S. Neergaard-Nielsen}%
    \affiliation{Center for Macroscopic Quantum States (bigQ), Department of Physics, Technical University of Denmark, 2800 Kongens Lyngby, Denmark}
\author{Ulrik L. Andersen}%
    \affiliation{Center for Macroscopic Quantum States (bigQ), Department of Physics, Technical University of Denmark, 2800 Kongens Lyngby, Denmark}
\date{\today}

\input{abstract}

\maketitle

\input{mainText}

\input{appendix}

\bibliography{bibliography.bib}

\end{document}

%% file: abstract.tex
\begin{abstract}
We analyze an all-optical setup which enables Bell-inequality violation over long distances by exploiting probabilistic entanglement swapping. The setup involves only two-mode squeezers, displacements, beamsplitters, and on/off detectors. We analyze a scenario with dichotomic inputs and outputs, and check the robustness of the Bell inequality violation for up to 6 parties, with respect to phase-, amplitude-, and dark-count noise, as well as loss.
\end{abstract}

%% file: mainText.tex
%Discovered by Boole in his work on probability theory \citep{Pitowsky:1994,Bancal:2010}, Bell inequalities constitute conditions on probabilities that result from an unknown local realist description. However, it was discovered by Bell that such inequalities could be violated within the framework of quantum mechanics, indicating that quantum mechanics cannot be recast as a local realist theory \citep{Bell}.
\section{Introduction}
\begin{center}
	\begin{minipage}{7cm}
	These ... may be termed conditions of possible experience. When satisfied they indicate that the data \textit{may} have, when not satisfied they indicate that the data \textit{cannot} have resulted from an actual observation.\\
	George Boole [1862]\\
\end{minipage}
\end{center}
As pointed out already by Boole in his work on probability theory, logical relations between observable events imply inequalities for the probabilities of their occurrence \citep{Pitowsky:1994,Bancal:2010}. Bell later demonstrated that the inequalities implied by a local realist description of nature can be violated within quantum mechanics \citep{Bell:1964}, implying that quantum mechanics cannot be recast as a local realist theory. Subsequent experimental investigations by Clauser, Aspect and their collaborators \citep{Aspect:1982a,Aspect:1982b,Freedman:1972} confirmed the nonlocal predictions of quantum mechanics, and nonlocality gradually became accepted as an aspect of nature. These early experiments were however not loophole-free, and while loophole-free violations have since been realised \cite{Hensen:2015,Shalm:2015,Giustina:2015}, it still remains experimentally challenging.\\

Loopholes constitute ways in which nature, or an eavesdropper, can arrange experimental outcomes, such that an experiment appears nonlocal, while in reality it is not. The detection loophole is relevant when inconclusive measurements are discarded from the experimental data \citep{Brunner:2014}. Such inconclusive measurements typically occur due to losses during transmission of the particles, or non-unit efficiency of the detectors. It has been demonstrated that discarding inconclusive measurement rounds renders it possible to violate a Bell inequality using classical optics \citep{Ilja:2011}. The locality loophole is present if measurements are performed such that a sub-luminal signal can transfer information between measurement stations during a measurement sequence. Such a sequence includes the act of choosing a measurement basis, and performing the measurement in this basis. The locality loophole can be closed by separating the measurement stations and keeping the duration of the measurement sequence short. However, this separation tends to induce losses and noise in the state shared by the participants of the experiment, and these losses tend to make the shared quantum state local, i.e. it cannot be used to demonstrate a Bell inequality violation.\\

In spite of these difficulties, the utilization of nonlocality is now moving from fundamental science towards practical applications, where the provable nonlocality of a quantum state is used in device-independent protocols to certify the security of a cryptographic key \citep{Pironio:2009,Acin:2006}. Crucial to the realization of device-independent quantum key distribution is the ability to close relevant loopholes, and to demonstrate the violation of Bell inequalities across distances relevant for telecommunication.\\
\begin{figure*}
	\centering		
	\includegraphics[width=1.9\columnwidth]{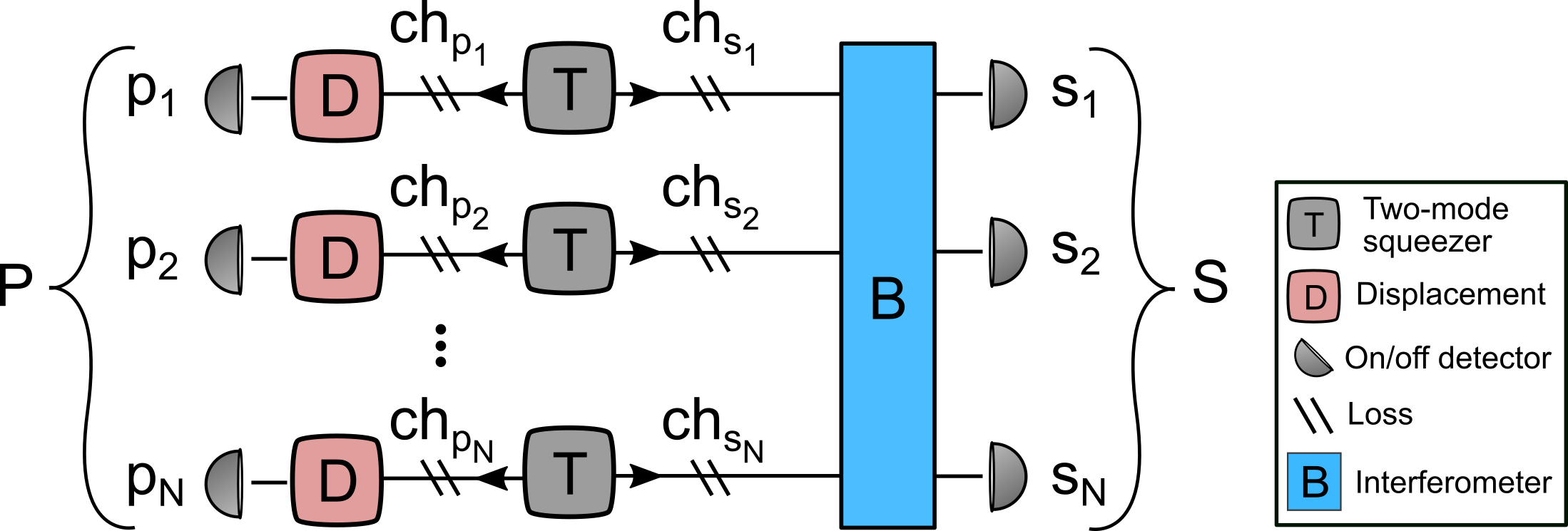}
		\caption{\textit{Sketch of the analysed setup with N parties. The left-going modes are labelled $p_n$ and the right-going modes are labelled $s_n$. A detector associated with a mode is given the same label as that mode. The measurement performed by the detectors in $S$ effectively swaps the N entangled states from the two-mode squeezers into an N-mode entangled state. ch abbreviates channel.}}
		\label{intro:fig:setup}
\end{figure*}

In this work we propose an experiment capable of violating a Bell inequality when the parties are separated by channels of low transmission. Our experiment is designed to be capable of closing the detection and locality loophole, and invokes only standard quantum optics tools, such as two-mode squeezers, displacements, and click detectors (on/off detectors). A sketch of the setup with N parties is shown in Fig. \ref{intro:fig:setup}. The proposed experiment is inspired by the setup in \citep{Brask:2012}, in which displacement-based measurements are used to demonstrate a Bell inequality violation. Two-mode squeezers (T) generate weakly squeezed two-mode squeezed vacuum states with half of each state sent a short distance to an on/off detector, and the other half sent to an interferometer B. The left-going modes in Fig. \ref{intro:fig:setup} are labelled $p_n$ and the right-going modes are labelled $s_n$, we group them into two sets $P=\{p_1,p_2,\ldots,p_N\}$ and $S=\{s_1,s_2,\ldots,s_N\}$. We use the same label for a mode and the corresponding detector. The interferometer B mixes the modes $S$, so that a photon arriving at one of the input ports of B, has an equal probability of triggering each of the detectors in $S$. We then require that only detector $s_N$ clicks, and that the remaining detectors in $S$ do not click, similar to an event-ready scheme \cite{Zukowski:1993}. Following this post-selection, the measurement outcomes for the detectors in $P$ are approximately the same as if the parties shared the single-photon state $\frac{1}{\sqrt{N}}(\ket{1,0,\ldots,0}+\ket{0,1,\ldots,0}+\ket{0,0,\ldots,1})$, in the limit of low squeezing. The nonlocality of this single-photon state was already analysed in \citep{Brask:2013,Laghaout:2011,Chaves:2011}, and we expect to see similar results for the approximate single-photon state analysed in this work.\\

Each detector in $P$ is considered as a party, with the possible measurement outcomes, click or no click, corresponding to whether any light arrives at the detector or not. Prior to each detector, either of two different displacements (D in Fig. \ref{intro:fig:setup}) is applied to the field. These displacements make up the two different measurement settings. We write the displacement applied on mode $p \in P$ as $X_p^{(n_p)} = \begin{pmatrix}  x_p^{(n_p)} & y_p^{(n_p)}  \end{pmatrix}^T$, with $n_p \in (0,1)$ labelling which of two possible displacements is implemented (measurement setting). We assume that all parties are choosing between the same two displacements, when the phases of the N two-mode squeezers are the same. This assumption is invoked to simplify our analysis, and we found no advantage when deviating from it. The displacement operator for mode $p$ is defined as,
\begin{align}
	D_p \left( X_p^{(n_p)} \right) = \exp\left[ i (\hat{q}_p \hspace{0.1cm} y_p^{(n_p)} - \hat{p}_p \hspace{0.1cm} x_p^{(n_p)}) \right] ,
\end{align}
where $\hat{q}_p$ and $\hat{p}_p$ are the quadrature operators for mode $p$. We follow the convention $[\hat{q}_k,\hat{p}_l]=2i\delta_{kl}$. From the quadrature operators we obtain the annihilation operator, $\hat{a}_p = \frac{1}{2} \left( \hat{q}_p + i\hat{p}_p \right) $. The coherent state generated by the displacement $X_p^{(n_p)}$, i.e. the state, $\ket{X_p^{(n_p)}} = D_p\left( X_p^{(n_p)} \right) \ket{0}$, is centred on the coordinates $\begin{pmatrix} q_p & p_p \end{pmatrix} =\begin{pmatrix} 2 x_p^{(n_p)} & 2 y_p^{(n_p)} \end{pmatrix}$ in phase space. We associate a click at a detector with the value 1, and no click with the value -1. The observable associated with detector $p$ is then given by,
\begin{align}
	M_p &= (I_p - \ket{0}_p \hspace{-0.09cm} \bra{0}) - \ket{0}_p \hspace{-0.09cm} \bra{0} \\
		&= I_p - 2 \ket{0}_p \hspace{-0.09cm} \bra{0} ,
\end{align}
where $I_p$ is the identity operator associated with mode $p$. We may transfer the displacement applied prior to detector $p$ onto the observable to obtain,
\begin{align}
	M_p^{(n_p)} = I_p - 2 \ket{-X_p^{(n_p)}}_p \hspace{-0.09cm} \bra{-X_p^{(n_p)}} .
	\label{Intro:Eq:Measurement}
\end{align}
We attempt to violate the W$^3$ZB (Werner-Wolf-Weinfurter-\.{Z}ukowski-Brukner) inequality \cite{Werner:2001,Weinfurter:2001,Zukowski:2001},
\begin{align}
	2^{-N} \sum_b \left\vert \sum_n (-1)^{\langle b,n \rangle} \langle M^{(n)} \rangle \right\vert \leq 1 .
	\label{Intro:Eq:W3ZB}
\end{align}
$b$ and $n$ are binary lists of length $N$, and the sums run over all possible binary lists. $\langle b,n \rangle$ is the dot product between $b$ and $n$. The entries of $n$ label the measurement settings of the involved parties. $\langle M^{(n)} \rangle$ is the correlator given by the product $\langle M^{(n)} \rangle = \langle \prod_p M_p^{(n_p)} \rangle$. We will refer to the left side of Eq. \ref{Intro:Eq:W3ZB} as the Bell value of the W$^3$ZB inequality. The maximal violation of the W$^3$ZB inequality increases with the number of parties \cite{Werner2:2001}. We therefore expect that when some loss and noise does not scale with the number of parties, then a violation of a W$^3$ZB inequality with more parties is more robust against this loss and noise, as compared to a W$^3$ZB inequality with fewer parties. \\
\begin{figure*}
	\centering
		\includegraphics[width=1.9\columnwidth]{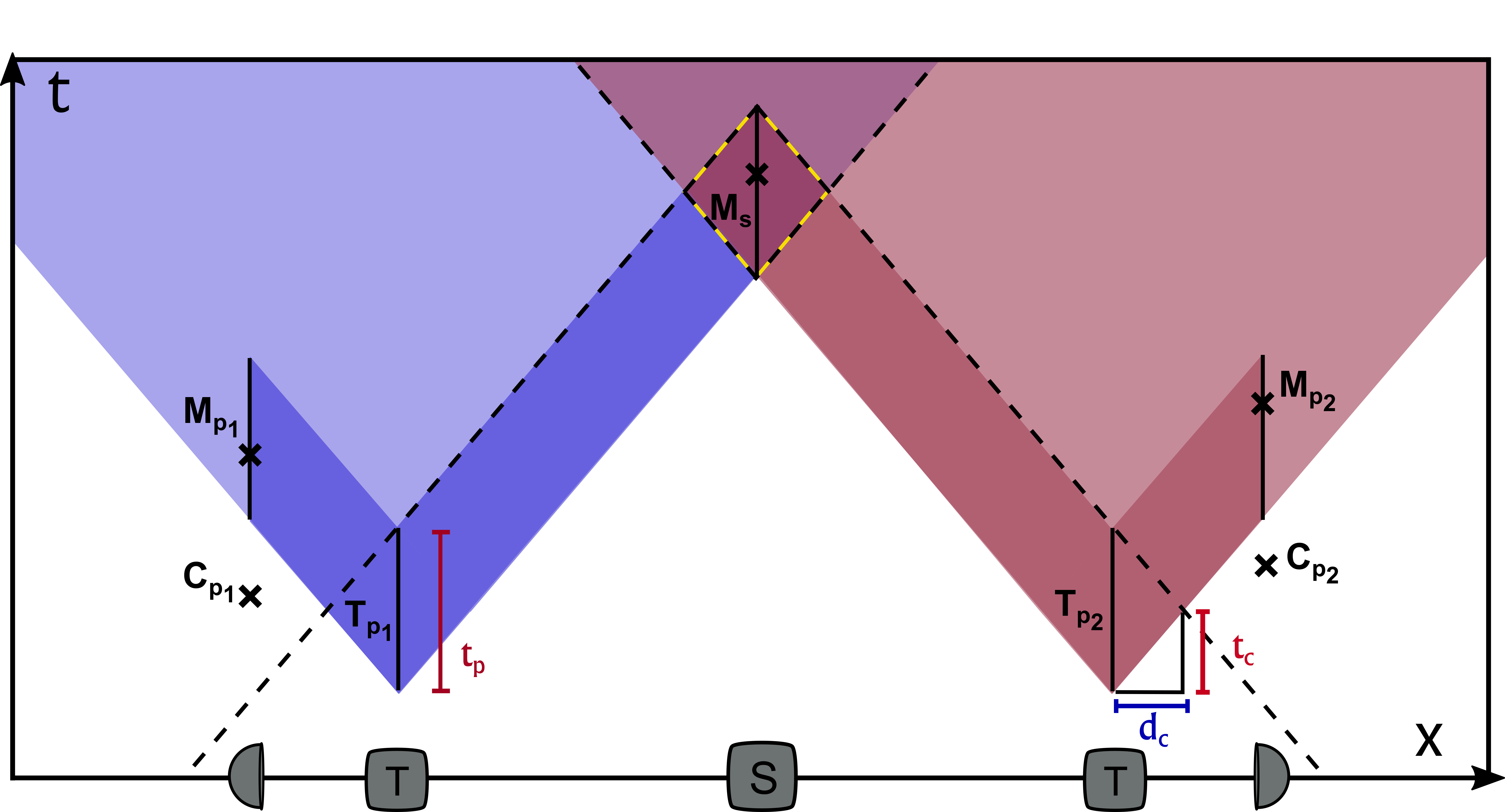}
		\caption{\textit{Space-time diagram of a loophole-free experiment with two parties, showing the space-time ordering of important events (marked by $\times$). The events T$_{p_1}$ and T$_{p_2}$ correspond to the generation of two-mode squeezed vacuum. C$_{p_1}$ and C$_{p_2}$ are the events where $p_1$ and $p_2$ decide their measurement settings. M$_{p_1}$ and M$_{p_2}$ correspond to events where $p_1$ and $p_2$ measure. M$_s$ correspond to the event where $s_1$ and $s_2$ measure. At the bottom we sketch the experimental setup (compare with Fig. \ref{intro:fig:setup}), where S corresponds to the swap following the interferometer B.}}
		\label{intro:fig:spacetime}
\end{figure*}

To close both the locality and detection loophole with two parties, $p_1$ and $p_2$, we require that the events of the experiment are positioned as shown in the space-time diagram in Fig. \ref{intro:fig:spacetime}.  The events T$_{p_1}$ and T$_{p_2}$ correspond to the generation of two-mode squeezed vacuum. These events occur along a temporal (vertical) line, since the light emitted from the source has a finite duration $t_p$. For this reason there exists at each position $x$ a duration of time where we expect the light to arrive with very high probability, this is marked with a darker shaded area. The measurements by $p_1$ and $p_2$ are labelled M$_{p_1}$ and M$_{p_2}$ respectively. M$_s$ correspond to the event where $s_1$ and $s_2$ measure. The choosing of measurement setting is labelled C$_{p_1}$ and C$_{p_2}$. The measurements M$_{p_1}$ and M$_{p_2}$ collapse the temporal width of the pulses, as illustrated in the figure by an $\times$. The swap M$_s$ occurs with very high probability along the vertical black line inside the central black and yellow dashed diamond. The backwards light cone for a swapping event will then typically be bounded by the dashed backwards light cone.\\

To close the detection loophole, $p_1$ and $p_2$ must choose their measurement settings at a time and place such that information about their choices cannot influence the swapping measurement M$_s$ via a sub-luminal signal. If the experiment is executed in this way, then we anticipate that an eavesdropper cannot tamper with the swap to falsify nonlocal correlations \cite{Bacciagaluppi:2021}. Most swapping events will obey this requirement if C$_{p_1}$ and C$_{p_2}$ are outside the dotted backward time cone shown in Fig. \ref{intro:fig:spacetime}. The critical distance $d_c$, which is the characteristic distance the event C$_{p_2}$ must be separated from the two-mode squeezer T$_{p_2}$, can be found by geometric arguments as $d_c=(1/2) c t_p$, and is associated with a waiting time $t_c = (1/2) t_p$. Ideally $p_2$ could make her choice of measurement setting at a distance $d_c$ from T$_{p_2}$, at a time $t_c$ after the light started to be emitted from the squeezer. Then her choice would most likely not be able to influence the swap M$_s$, while at the same time ensuring that the light pulse has not passed by her yet.\\

The experimental constraints discussed above generalize to the scenario where N parties attempt to obtain a Bell inequality violation, while closing the detection and locality loophole. That is, the parties should ensure that the events C$_{p_n}$ are outside the backward timecone for the swapping event M$_s$. However, one should also ensure that the parties are sufficiently distant from each other, so that information on the choice of measurement setting and outcome cannot travel between parties during a measurement sequence.

\section{Model}
We now give an outline of how we model the optical field, and how we include experimental imperfections in our analysis. A full description can be found in appendix A1.
The fields generated by the two-mode squeezers are distributed in time and space according to some mode functions \cite{Agarwal:2013}. The amplitudes of these modes are quantum uncertain with Gaussian statistics described by a covariance matrix $\sigma$ with elements $\sigma_{kl}=1/2 \langle \{Q_k,Q_l \} \rangle - \langle Q_k \rangle \langle Q_l \rangle$, where $\{.,.\}$ denotes the anti-commutator and $Q=Q_P \oplus Q_S$, where $Q_P = \bigoplus_{p \in P} \begin{pmatrix} \hat{q}_{p} & \hat{p}_{p} \end{pmatrix}$ with $Q_S = \bigoplus_{s \in S} \begin{pmatrix} \hat{q}_{s} & \hat{p}_{s} \end{pmatrix}$ \cite{Weedbrook:2012}. The corresponding density matrix, also describing the statistics of the field, is denoted $\rho$. We denote the squeezing parameter of the N squeezers as $r$ and introduce the symbols, $a=\sinh(2r)$ and $v=\cosh(2r)$. The covariance matrix of the 2N modes can be written as,
\begin{align}
	\sigma = \begin{pmatrix}
	v \mathbf{I} & \mathbf{R}_\phi \\
	\mathbf{R}_\phi & v\mathbf{I}
	\end{pmatrix} ,
	\label{Model:Eq:Cov}
\end{align}
where $\mathbf{I}$ is the identity matrix of dimension 2N and $\mathbf{R}_\phi$ is the block diagonal matrix,
\begin{align}
	\mathbf{R}_\phi = \bigoplus_{p} \begin{pmatrix}
	a \cos(\phi_p) & -a\sin(\phi_p) \\
-a\sin(\phi_p) & -a \cos(\phi_p)
	\end{pmatrix} , 
\end{align}
where $\phi_p$ is the phase angle of the squeezer for party $p$. The expectation value of the field amplitude is assumed zero. The Wigner characteristic function corresponding to $\rho$ is given by $\chi_\rho(\Lambda) = \exp[-(1/2) \Lambda^T \Omega \sigma \Omega^T \Lambda]$ where $\Lambda$ is a vector of conjugate quadratures (the Fourier transform dual to the quadratures)  for the modes $P$ and $S$, i.e. $\Lambda = \Lambda_P \oplus \Lambda_S$, where $\Lambda_P = \bigoplus_{p \in P} \Lambda_p$ and $\Lambda_S = \bigoplus_{s \in S} \Lambda_s$. The conjugate quadratures for mode $k$ is a vector $\Lambda_k = \begin{pmatrix} \lambda_{kx} &\lambda_{ky} \end{pmatrix}^T$. We have also introduced the symplectic form $\Omega = \bigoplus_{k = 1}^{2N} \omega$, where $\omega$ is the antisymmetric matrix,
\begin{align}
	\omega = \begin{pmatrix} 0 & 1 \\ -1 & 0 \end{pmatrix} .
\end{align}\\
The modes $S$ are then mixed on the interferometer B, and we assume that the corresponding mode functions are identical and have a high overlap at the beamsplitters making up the interferometer. Let $\hat{a}_s$ be the amplitude operator for a mode $s \in S$, the interferometer B is assumed to generate the Bogoliubov transformation,
\small
\begin{align}
	\begin{pmatrix}
		\hat{a}_{s_1} \\
		\hat{a}_{s_2} \\
		\hat{a}_{s_3} \\
		\vdots \\
		\hat{a}_{s_N}
	\end{pmatrix}
	 \rightarrow \frac{1}{\sqrt{N}}
	 \begin{pmatrix}
	 		1 & e^{i \frac{2 \pi}{N} } & e^{i 2 \frac{2 \pi}{N} } & \ldots & e^{i (N-1) \frac{2 \pi}{N} } \\ 
	 		1 & e^{i 2 \frac{2 \pi}{N} } & e^{i 4 \frac{2 \pi}{N} } & \ldots & e^{i 2(N-1) \frac{2 \pi}{N} } \\
	 		1 & e^{i 3 \frac{2 \pi}{N} } & e^{i 6 \frac{2 \pi}{N} } & \ldots & e^{i 3(N-1) \frac{2 \pi}{N} } \\
	 		\vdots & \vdots & \vdots & \ddots & \vdots \\
1 & 1 & 1 & \ldots & 1
	 \end{pmatrix}
	\begin{pmatrix}
		\hat{a}_{s_1} \\
		\hat{a}_{s_2} \\
		\hat{a}_{s_3} \\
		\vdots \\
		\hat{a}_{s_N}
	\end{pmatrix}	
	\label{model:eq:bogo}
\end{align}
\normalsize
We condition the state on obtaining a click at detector $s_N$ and no clicks at the remaining detectors, thereby heralding the conditional state $\rho_c$ of modes $P$. The projector corresponding to this event is $\hat{\Pi}_c= \left( \prod_{s \in \bar{S}} \ket{0}_s \hspace{-0.09cm} \bra{0} \right) (I_{s_N}-\ket{0}_{s_N} \hspace{-0.09cm} \bra{0})$, where $\bar{S}$ is the set $\bar{S} = S \backslash \{s_N \}$. The conditional state is obtained as $\rho_c =  \operatorname{Tr}_{S} [ \rho \hat{\Pi}_c ]  / P(C)$, where $P(C)$ is the normalization, i.e. the probability of obtaining the measurement outcomes heralding a successful swap. $\hat{\Pi}_c$ has the characteristic function,
 \begin{align}
	&\chi_c(\Lambda_S) = \operatorname{Tr} \left[ \hat{\Pi}_c D_{S}(\Lambda_{S}) \right] \nonumber \\ &=  E(\Lambda_{\bar{S}}) \cdot \left( \pi \delta^{(2)}(\Lambda_{s_N}) - E(\Lambda_{s_N}) \right) ,
\end{align}
where
\begin{align}
	E (\Lambda_j) = \exp\left[ -\frac{1}{2} \Lambda_j^T \Lambda_j \right], 
\end{align}
and $\delta^{(2)}(\Lambda_j)$ is a delta function. We obtain the characteristic function of the conditional state through integration,
\begin{align}
	\chi_{\rho_c} (\Lambda_P) = \frac{1}{\pi^N P(C)} \int_{\mathbb{R}^{2N}} \chi_{\rho}(\Lambda) \chi_c(-\Lambda_{S}) d^{2N} \Lambda_{S} .
\end{align}
We then compute the Bell value of the W$^3$ZB inequality by evaluating the expectation values $\langle M^{(n)} \rangle = \langle \prod_{p \in P} M_p^{(n_p)} \rangle$, for each setting $n$. This is done via the integral \cite{Ferraro:2005},
\begin{align}
	& \left\langle \prod_{p \in P} M_p^{(n_p)} \right\rangle = \operatorname{Tr}\left\{\rho_c \prod_{p \in P} M_p^{\left(n_p\right)}\right\} \nonumber \\
	 &=\frac{1}{\pi^{N}} \int_{\mathbb{R}^{2N}} \chi_{\rho_c}(-\Lambda_P) \chi_M\left(\Lambda_P, X_{P}\right) d^{2 N} \Lambda_P ,
\end{align} 
where $\chi_M\left(\Lambda_P, X_{P}\right)$ is the characteristic function associated with the observable $\prod_{p \in P} M_p^{(n_p)}$. $X_P$ is a vector of the displacements applied prior to the detectors, $X_P = \bigoplus_{p \in P} X_p^{(n_p)}$. A closed form expression for $\langle \prod_{p \in P} M_p^{(n_p)} \rangle$ can be found in appendix A1.
\subsection*{Noise model}

We now outline how we describe noise relevant to the experiment. We will include dark-counts in the detectors, loss in the channels, phase noise in the channels and measurements, and finally, amplitude noise in the measurements. Amplitude and phase noise during measurement are expected to arise if imperfect displacements are applied.

We include dark-counts in our measurement model by adding a noise term to the observable. Given that $p_d$ is the probability of getting a dark-count during the measurement interval, then we measure the observable,
\begin{align}
M_p^{\left(n_p\right)}&=\left(1-p_d\right)\left[I_p-2\left|-X_p^{\left(n_p\right)}\right\rangle_p \hspace{-0.09cm} \left\langle-X_p^{\left(n_p\right)}\right|\right]+p_d I_p \nonumber \\
&= I_p-2(1-p_d)\left|-X_p^{(n_p)}\right\rangle_p \hspace{-0.09cm} \left\langle-X_p^{(s_p)}\right| .
\end{align}
If the detectors in $S$ are triggered by a dark-count with probability $p_d$, then the swap results in the transformation (see appendix A1),
\begin{align}
	&\rho \rightarrow \rho_c = \frac{1}{P(C)} \operatorname{Tr}_{S} \left[ \rho \widetilde{\Pi}_c \right] .
\end{align}
We have introduced the operator $\widetilde{\Pi}_c$,
\begin{align}
&\widetilde{\Pi}_c = \nonumber \\ & (1-p_d)^{N-1} &\left( \prod_{s \in \bar{S}} \ket{0}_s \hspace{-0.09cm} \bra{0} \right) \cdot \left( I_{s_N}-(1-p_d)\ket{0}_{s_N} \hspace{-0.09cm} \bra{0} \right) .
\end{align}
Given that channel ch$_{p_n}$ has transmission $\eta_{p_n}$ and channel ch$_{s_k}$ has transmission $\eta_{s_k}$, we model loss by a Gaussian map acting on the covariance matrix $\sigma$ as \cite{Ferraro:2005},
\begin{align}
\sigma \rightarrow G_\eta^{1 / 2} \sigma G_\eta^{1 / 2}+\left(I-G_\eta\right) ,
\label{model:eq:lossMap}
\end{align}
with the diagonal matrix $G_\eta= G_{\eta_P} \oplus G_{\eta_S}$, where $G_{\eta_P} = \operatorname{Diag} \left[ \bigoplus_{p \in P} \begin{pmatrix} \eta_p  & \eta_p \end{pmatrix} \right]$ and $G_{\eta_S} = \operatorname{Diag} \left[  \bigoplus_{s \in S} \begin{pmatrix} \eta_s & \eta_s \end{pmatrix} \right]$. We will assume that $\eta_{p_n}$ equals $\eta_P$, i.e. the channels ch$_{\texttt{P}} = \{$ch$_{p_1}$,ch$_{p_2}$,$\ldots$,ch$_{p_N}\}$ have the same transmission. Likewise we assume that $\eta_{s_k}$ equals $\eta_S$. $\eta_d$ is the efficiency of a detector, and $1-\eta_d$ is the loss of the detector. Given that $\eta_d$ is the same for all detectors in $S$, detector loss can then be commuted through B and absorbed into the transmission of the channels ch$_{\texttt{S}} = \{$ch$_{s_1}$,ch$_{s_2}$,$\ldots$,ch$_{s_N}\}$. Likewise, the detector loss in $P$ can be shifted to be prior to the displacements, if we attenuate the magnitude of the displacements by the factor $\sqrt{\eta_d}$. \\
%So if the transmission of channels ch$_P$ is $\eta_p$, and the efficiency of each detector in P is $\eta_d$, then we can set the efficiency of the detectors to unity, if we set the transmission of channels ch$_P$ to $\eta_p \eta_d$. Likewise, we can set the efficiency of the detectors at S to unity, if we set the transmissions of channels ch$_S$ to $\eta_s \eta_d$. \\
A phase perturbation of the state $\rho$, e.g. caused by environmental disturbance, can be modelled as a stochastic rotation in phase space,
\begin{align}
\rho=\int d^{N} \boldsymbol{\theta} P(\boldsymbol{\theta}) R(\boldsymbol{\theta}) \rho_0 R(-\boldsymbol{\theta}) ,
\label{model:equation:phasenoise}
\end{align}
where $\rho_0$ is the unperturbed state, and $\boldsymbol{\theta}$ is a vector of stochastic rotation angles $\theta_{p}$ for $p \in P$, each being a perturbation on the phase of the corresponding mode. Note that phase noise acting on channels ch$_{\texttt{S}}$ is shifted to act on channels  ch$_{\texttt{P}}$ instead. $R(\boldsymbol{\theta})$ is the rotation operator $R(\boldsymbol{\theta})=\prod_{p \in P} R_p\left(\theta_p\right)$. $R_{p_n}\left(\theta_{p_n}\right)$ is applied just prior to the displacement operation on mode $p_n$, and includes phase noise resulting from propagation in the channels ch$_{\texttt{p}_n}$ and ch$_{\texttt{s}_n}$, and also the phase noise in the subsequent displacement operation. We make the assumption that the angles $\boldsymbol{\theta}$ are uncorrelated, and model the probability density $P(\boldsymbol{\theta})$ as a product of normal distributions for each angle $\theta_{p}$. The variance of $\theta_{p}$ is labelled as $V_\theta$, and is the same for all modes. The correlated phase noise resulting from the interferometer B cannot be entirely captured by this simple model, but we expect that our model is sufficiently close to reality to indicate the sensitivity of the experiment toward phase noise. We furthermore assume that the angles $\theta_{p}$ are small, allowing us to approximate the rotation of a coherent state by a small linear translation in phase space. \\
Amplitude noise arises from an imperfect displacement and is modelled similarly to phase noise, with the rotation operator in Eq. \ref{model:equation:phasenoise} replaced by a displacement operator. The stochastic displacement on mode $p$ is given relative to the displacement $X_p^{(n_p)}$ applied on mode $p$, i.e. for mode $p$ we obtain the stochastic displacement $r_p X_p^{(n_p)}$, where $r_p$ is referred to as the relative amplitude. We assume that the relative amplitudes $r_p$ are normal, independent and identically distributed, with variance $V_A$. A more detailed description of the noise model can be found in appendix A1.

\section{Results and Discussion}
\begin{figure}
	\begin{center}
		\includegraphics[width=1\columnwidth]{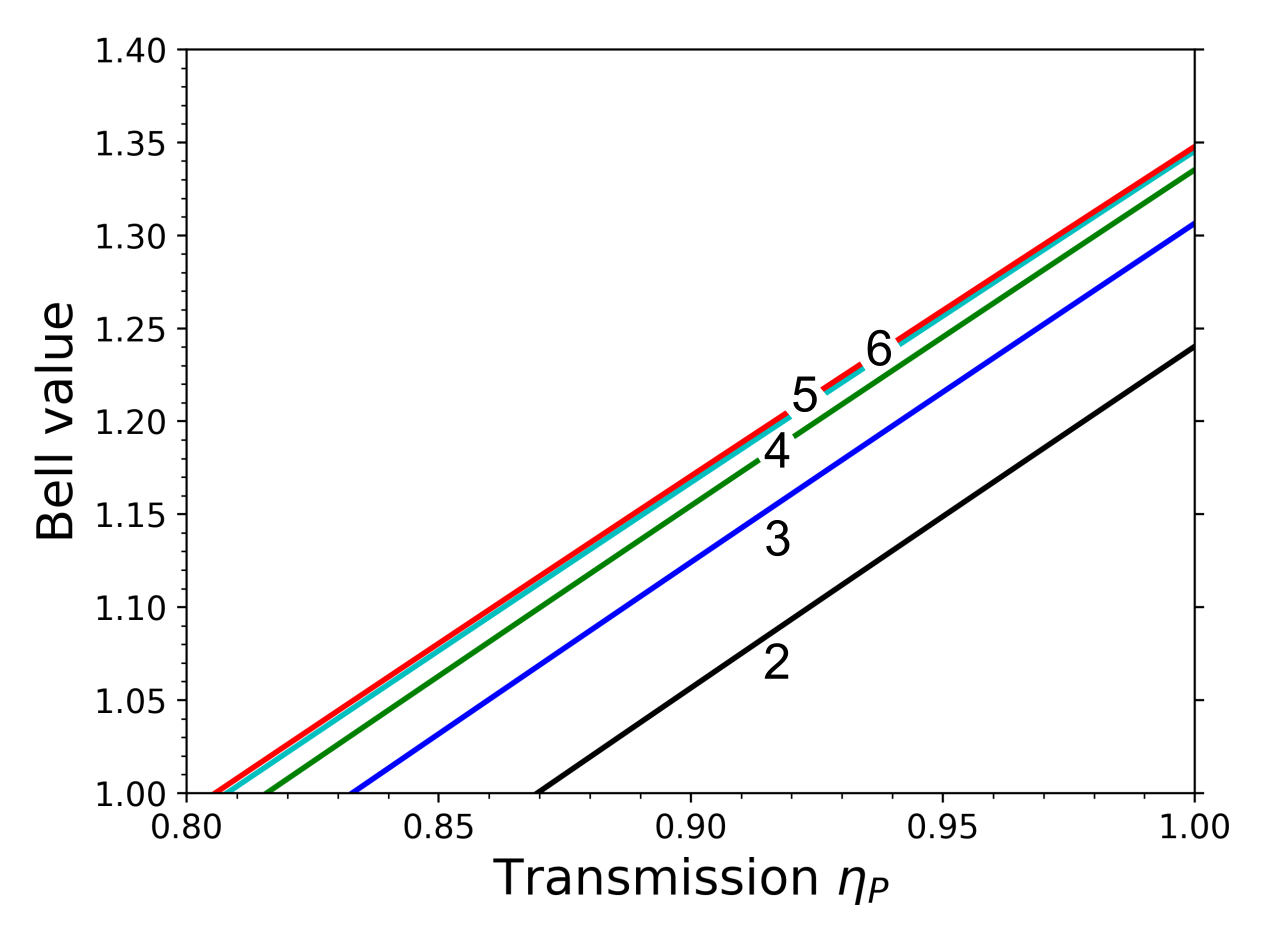}
		\caption{\textit{We plot how the Bell value of the W$^3$ZB inequality depends on the transmission of the channels ch$_\texttt{P}$, connecting the two-mode squeezers to the detectors in $P$. The annotation indicates the number of parties.}}
		\label{results:fig:etaP}
	\end{center}
\end{figure}

We compute Bell values under varying experimental conditions. In order to obtain realistic values we must include in the model reasonable experimental errors. We choose the noise parameters shown in Table \ref{results:table:noise}. Unless otherwise stated, these are the values used for the noise parameters throughout our analysis. E.g. if we vary $\eta_P$, as is done in Fig. \ref{results:fig:etaP}, then the remaining noise parameters are set at the values listed in Table \ref{results:table:noise}. \\
\begin{table}
\begin{center}
\begin{tabular}{c|c|c|c|c}
 $\eta_P$ & $\eta_S$ & $\sigma_A$ & $\sigma_\theta$ & $P_d$ \\
\hline $0.9$ & $0.2$ & $3 / 100$ & $100 \ \mathrm{mrad}$ & $1 / 10000$ \\
\end{tabular}
\caption{\textit{Standard settings for noise parameters. $\eta_P$ is the transmission of the channels ch$_\texttt{P}$ (ch$_{\texttt{p}_1}$, ch$_{\texttt{p}_2}$ etc.). $\eta_S$ is the transmission of channels ch$_\texttt{S}$. $\sigma_A$ is the standard deviation of the relative amplitude distribution $\left(\sigma_A^2=V_A\right)$. $\sigma_\theta$ is the standard deviation of the phase angle distribution $\left(\sigma_\theta^2=V_\theta\right)$. $P_d$ is the probability of getting a dark-count during the measurement interval (which is assumed to be $t_p$ in our analysis).}}
\label{results:table:noise}
\end{center}
\end{table}

\begin{figure}
	\begin{center}
		\includegraphics[width=1\columnwidth]{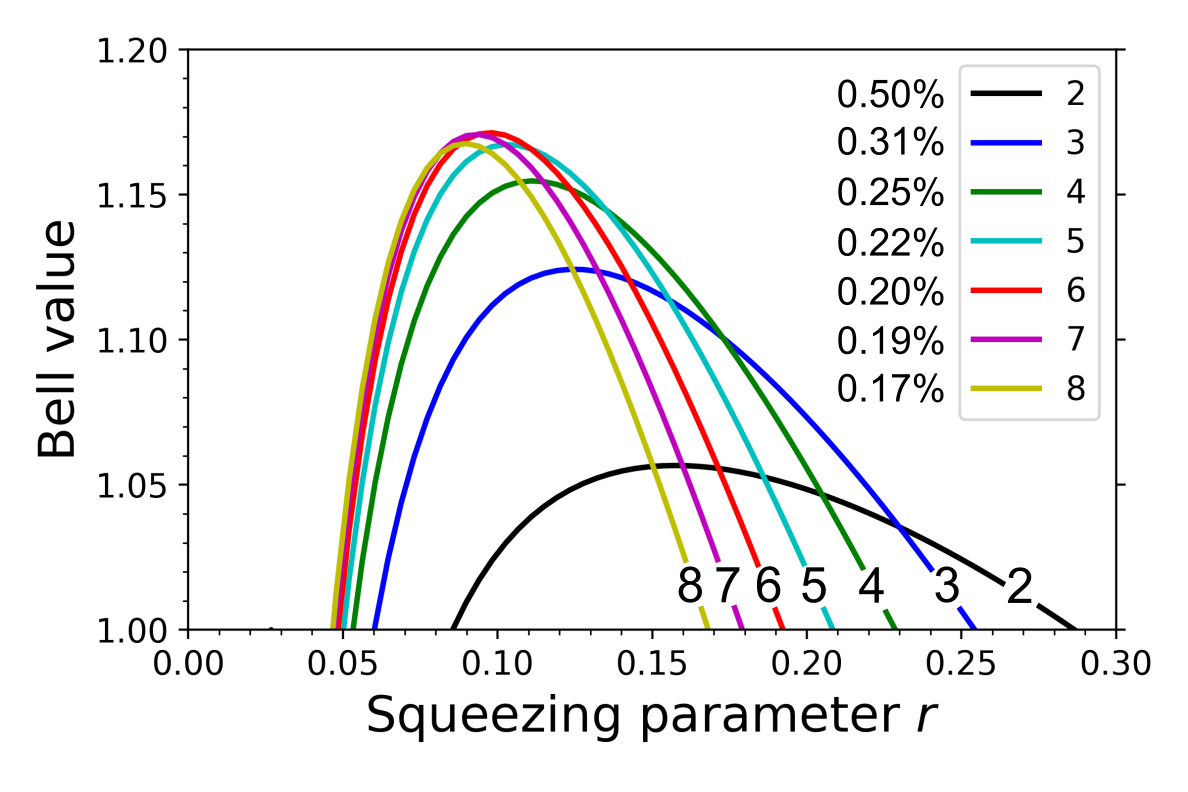}
		\caption{\textit{Bell value against the squeezing parameter r for different number of parties. The annotation and legend gives the number of parties. The Bell value is computed for the optimal measurement settings at the given value of $r$.  We observe a maximum in the Bell value at a particular squeezing. Next to the legend we list the probability $P(C)$ that an experiment succeeds with that number of parties, at the corresponding optimal value of $r$.}}
		\label{results:fig:squeezing}
	\end{center}
\end{figure}

We maximize the violation of the W$^3$ZB inequality in the squeezing parameter $r$. The Bell value as a function of $r$, for the optimal choice of measurement settings, is shown in Fig. \ref{results:fig:squeezing}. We clearly observe that there exists an optimal squeezing value for which the Bell value is maximized, and that the optimal squeezing depends on the number of parties. We also observe that the maximal Bell value increases for more parties, until 6 parties, at which point the maximal Bell value decreases for more parties.\\
While the correlations between all parties lead to a violation of the $\mathrm{W}^3 \mathrm{ZB}$ inequality at the optimal squeezing, we find that, for up to 4 parties, the marginal outcome probabilities describing any subgroup of parties are inside the Bell polytope, with the used measurement settings. This was evidenced by a linear program (see appendix A2), and indicates that in these cases nonlocality results from correlations between \textit{all} parties. An exception can occur for 5 parties if $\eta_P$ is above $97 \%$, and for 6 parties if $\eta_P$ is above $91 \%$, with the used measurement settings. In these cases a Bell inequality can be broken with a subgroup of 4 and 5 parties respectively.\\
% In order to prove nonlocal correlations between all parties in an experimental realization, it will suffice to verify that all subgroups satisfy the W$^3$ZB inequality \cite{Werner:2001}, and that relevant loopholes are closed.

\begin{figure}
	\begin{center}	\includegraphics[width=0.9\columnwidth]{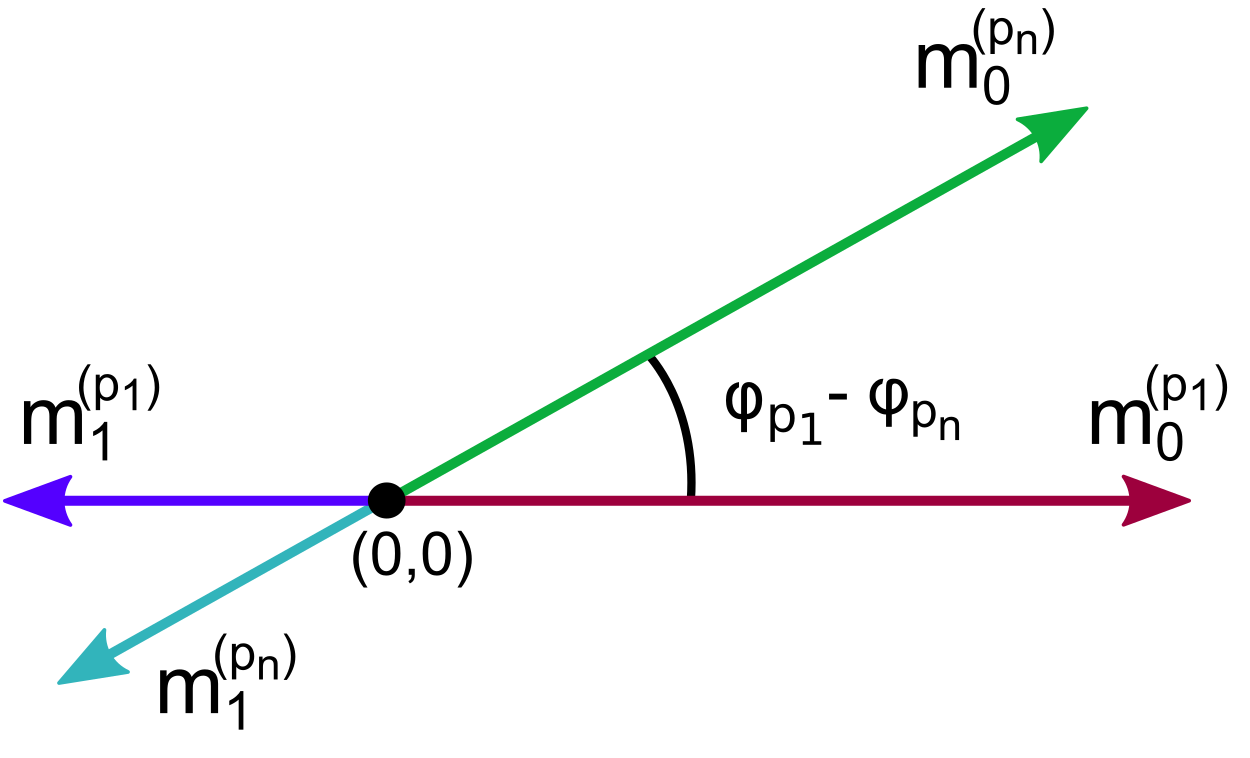}
		\caption{\textit{We show the orientation of the optimal displacements (measurement settings) that the parties $p_1$ and $p_n$ should use to obtain a maximal violation of the $W^3$ZB inequality. $m_0^{(p_1)}$ and $m_1^{(p_1)}$ are the displacements used by party $p_1$, whereas $m_0^{(p_n)}$ and $m_1^{(p_n)}$ are the displacements used by party $p_n$ (any party). $p_n$’s displacements should be at the angle $\phi_{p_1}-\phi_{p_n}$ relative to $p_1$'s displacements, where $\phi_{p_1}$ and $\phi_{p_n}$ are the phase angles of the two-mode squeezers belonging to $p_1$ and $p_n$. The magnitudes of $m_0^{(p_n)}$ and $m_1^{(p_n)}$ depend on the number of parties and are listed in Table \ref{Results:Table:amplitudes}.}}
		\label{results:fig:settings}
	\end{center}
\end{figure}

We find the optimal displacements (measurement settings), at the optimal squeezing, for which the violation is maximized. The optimal displacement for party $p_1$ and another party $p_n$, are shown in Fig. \ref{results:fig:settings}. The phase angles of the two-mode squeezers belonging to $p_1$ and $p_n$ respectively, are labelled as $\phi_{p_1}$ and $\phi_{p_n}$. $m_0^{(p_1)}$ and $m_1^{(p_1)}$ are the displacements used by party $p_1$, whereas $m_0^{(p_n)}$ and $m_1^{(p_n)}$ are the displacements used by party $p_n$. $m_0^{(p_1)}$ and $m_0^{(p_n)}$ have the same magnitude, but the displacements are directed along different quadrature axes at an angle $\phi_{p_1}-\phi_{p_n}$, likewise for $m_1^{(p_1)}$ and $m_1^{(p_n)}$. So the displacements used by a given party $p_n$ will be determined by the phase angle of their squeezer $\phi_{p_n}$. The magnitudes of $m_0^{(p_n)}$ and $m_1^{(p_n)}$ depend on the number of parties and are listed in Table \ref{Results:Table:amplitudes}. The overall orientation of the quadrature axes is arbitrary, i.e. we can freely rotate Fig. \ref{results:fig:settings}, as long as the angle between displacements remain unchanged. In this sense, the displacements used by party $p_1$ serve as a reference from which we can define the displacements to be used by the remaining parties. \\

We note that the optimum in squeezing, seen in Fig. \ref{results:fig:squeezing}, is the result of a competition between the dark-count rate and the multi-photon generation rate. A dark-count would render the measurements by the parties uncorrelated, thereby lowering the Bell value. This indicates that it is preferable to have high squeezing, so that photons from the optical field outnumber the dark-counts. However, the click detectors in $S$ cannot distinguish between 1 or more photons. Multi-photon emission from the two-mode squeezers therefore create mixedness in the conditional state generated by the swap, and this mixedness weakens the correlations between the measurement outcomes obtained by the parties. This mixedness can be avoided by lowering the degree of squeezing, so that on average less than one photon reaches the detectors in $S$. As a result, there is some amount of squeezing where the combined detrimental effect of dark-counts and multi-photon generation is minimized. As we increase the number of parties, the presence of dark-counts becomes more detrimental due to the increased number of detectors, and the lower probability of a successful swap $P(C)$. This is the cause of the decrease in maximal Bell value for 7 and 8 parties, as compared to the case with 6 parties.\\
\begin{table}[]
\begin{center}
\begin{tabular}{c|c|c}
No. of Parties & $m_0$   & $m_1$    \\ \hline
2 & 0.59 & -0.18 \\ \hline
3 & 0.47 & -0.20 \\ \hline
4 & 0.41 & -0.19 \\ \hline
5 & 0.37 & -0.18 \\ \hline
6 & 0.33 & -0.17
\end{tabular}
\caption{\textit{Magnitudes of the optimal displacements shown in Fig. \ref{results:fig:settings} for the optimal value of $r$. If the detector transmission is $\eta_d$, then the magnitudes should be multiplied by $1/\sqrt{\eta_d}$.}}
\label{Results:Table:amplitudes}
\end{center}
\end{table}

\begin{figure*}
\centering
		\includegraphics[width=2\columnwidth]{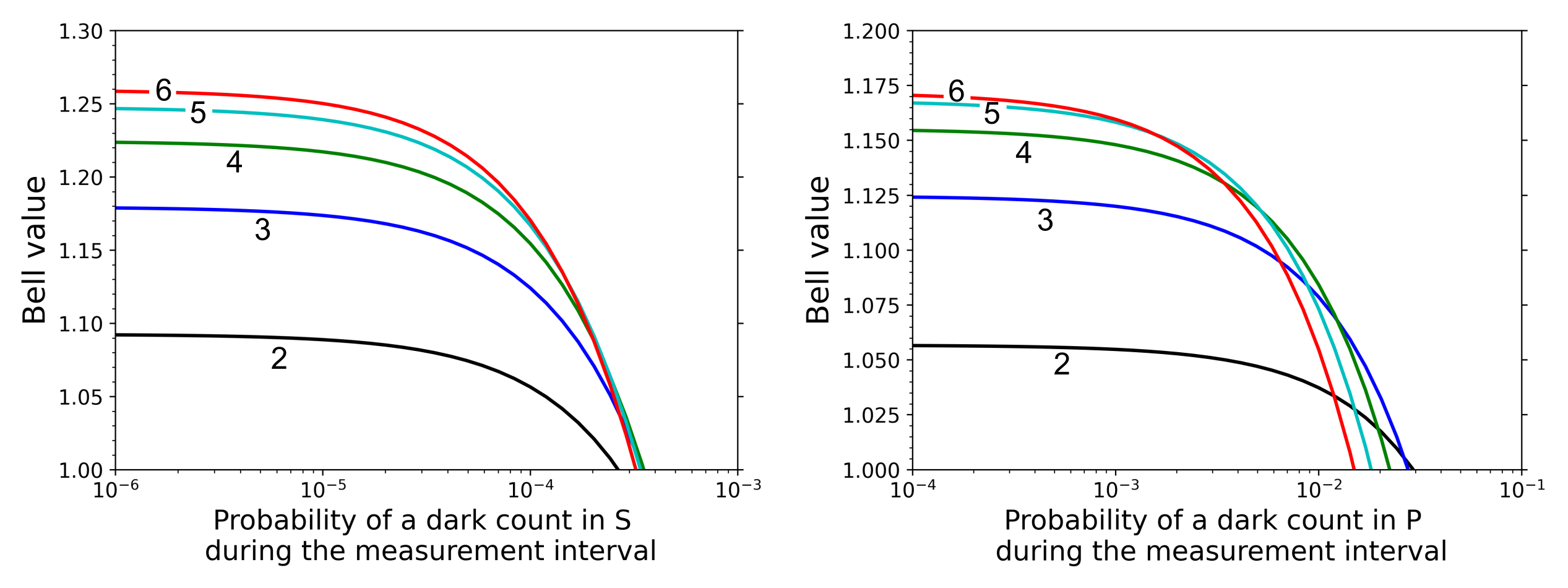}
		\caption{\textit{\textbf{Left}: Bell value of the W$^3$ZB inequality against the probability of a dark-count in $S$ during the measurement interval. The annotation indicates the number of parties. \textbf{Right}:  Bell value of the W$^3$ZB inequality against the probability of a dark-count in $P$ during the measurement interval.}}
		\label{results:fig:darkCounts}
\end{figure*}

We investigate the sensitivity of the experiment against a dark-count at a detector in $S$, the result can be seen in Fig. \ref{results:fig:darkCounts} (left) for different number of parties. A dark-count at detector $s_N$ would mistakenly herald nonlocal correlations between the detectors in $P$, when no such correlations actually exists. This erroneous heralding significantly lowers the calculated Bell value. The Bell value is found to rapidly decrease around $P_d \approx$ $0.02 \%$. At this point, the probability of getting a dark-count is no longer insignificant compared with the probability of generating the conditional state, which is in the range $0.2\%$ to $0.5\%$, depending on the number of parties (see Fig. \ref{results:fig:squeezing}). For the case of 2 parties, the decrease in Bell value proceeds a bit slower, however the lower initial Bell value (1.09) results in the curve reaching the classical limit of 1 at smaller dark-count probabilities. \\
We also analyse the robustness of the Bell inequality violation against dark-counts at the detectors in $P$. A plot of the Bell value against the probability of a dark-count in $P$ is shown in Fig. \ref{results:fig:darkCounts} (right), and clearly illustrates that the violation is highly robust against such a dark-count.\\

\begin{figure}
	\begin{center}
		\includegraphics[width=1\columnwidth]{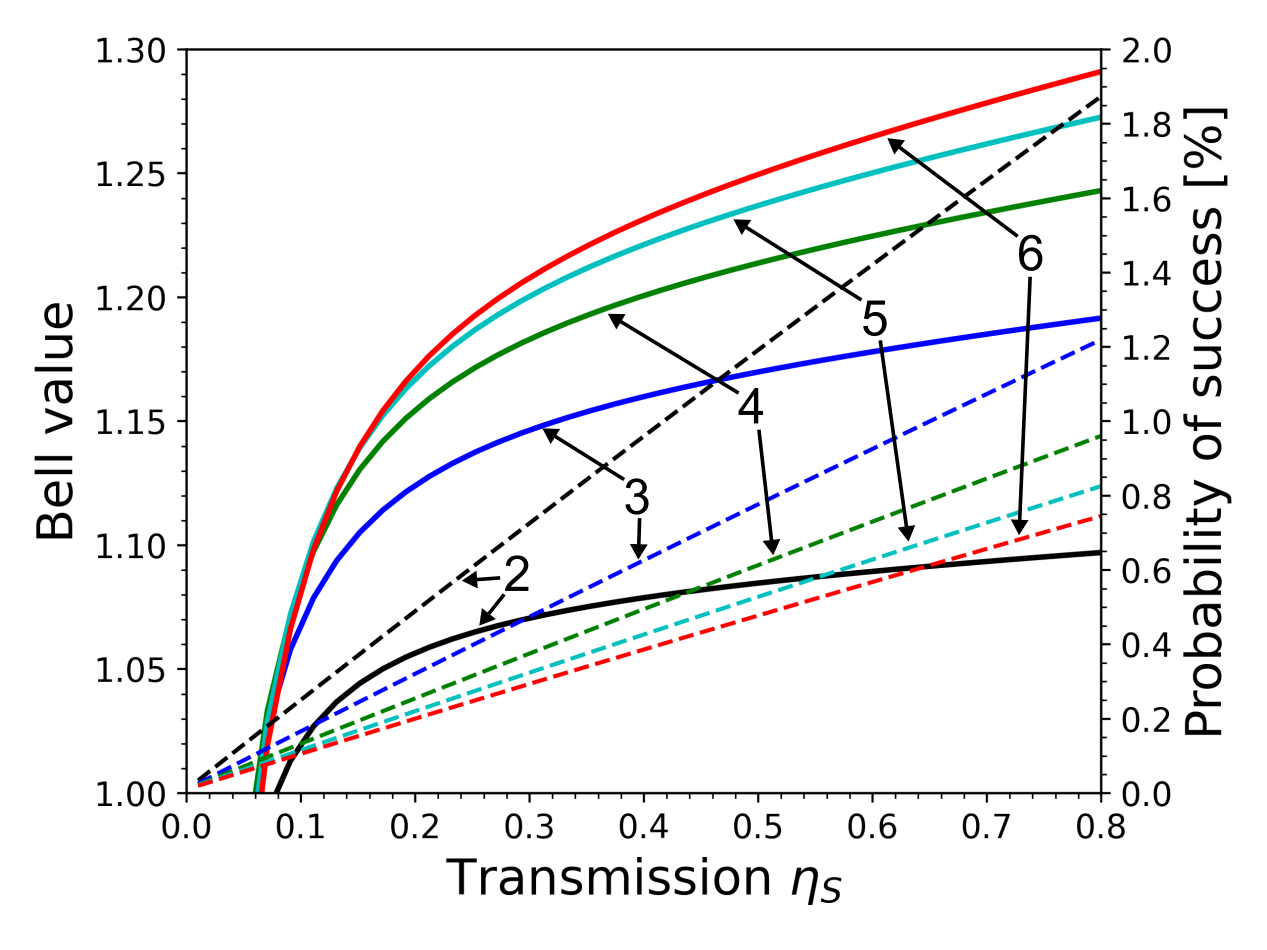}
		\caption{\textit{We plot how the Bell value of the W$^3$ZB inequality depends on the transmission of the channels ch$_S$, connecting the two-mode squeezers to the swapping detectors $S$. The annotation indicates the number of parties. The solid curves correspond to Bell values and match the left y-axis. The dashed curves are the corresponding probabilities of generating the conditional state, these drop as we lower the transmission $\eta_S$.}}
		\label{results:fig:etaS}
	\end{center}
\end{figure}

The impact of loss on the Bell value of the W$^3$ZB inequality is shown in Fig. \ref{results:fig:etaP} and Fig. \ref{results:fig:etaS}. In Fig. \ref{results:fig:etaP} we vary the transmission $\eta_P$, and show how the Bell value changes. The transmission at which the Bell value drop below one, lowers as we increase the number of parties. This indicates that a demonstration of nonlocality might be easier to realize when using more parties.
In Fig. \ref{results:fig:etaS} we show the dependence of the Bell value on the transmission $\eta_S$. We observe that the Bell value is only weakly dependent on this transmission until a critical point around a transmission of 10 \%. The probability $P(C)$ of successfully generating the conditional state, heralded by detector $s_N$ clicking and the remaining detectors in $S$ staying silent, is seen to drop linearly for decreasing transmission. If we assume a fiber loss of 0.3 dB/km, we find that a transmission of 10 \% corresponds to approximately 30 km. The maximal achievable separation between two parties will then be around 60 km.\\

\begin{figure*}
\centering
		\includegraphics[width=2\columnwidth]{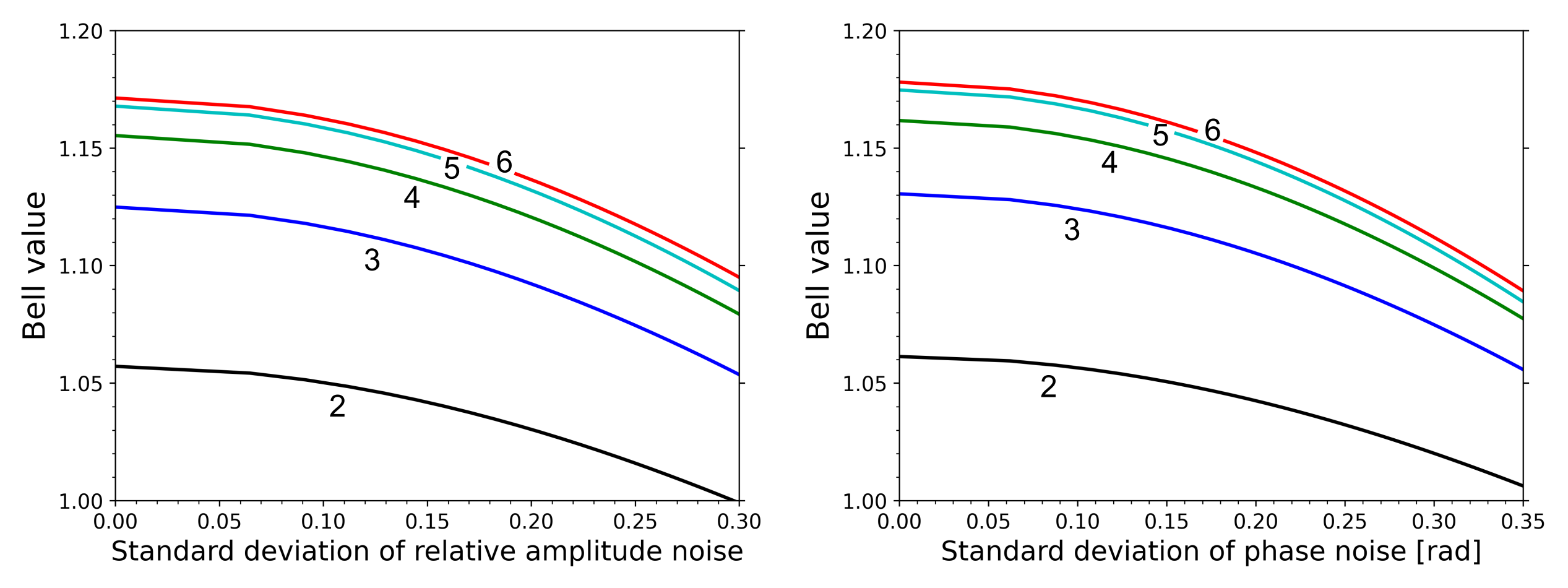}
		\caption{\textit{\textbf{Left}: We plot how the Bell value depends on amplitude noise ($\sigma_A$). The annotation indicates the number of parties. \textbf{Right}: We plot how the Bell value depends on the phase noise ($\sigma_\theta$).}}
		\label{results:fig:ampphs}
\end{figure*}

We then check the sensitivity of the experiment against phase and amplitude noise. The result is shown in Fig. \ref{results:fig:ampphs}. In Fig. \ref{results:fig:ampphs} (left) we plot the Bell value against the standard deviation of the relative amplitude distribution, $\sigma_A$. In Fig. \ref{results:fig:ampphs} (right) we plot the Bell value against the standard deviation of the phase distribution, $\sigma_\theta$. We observe that the Bell value is not very sensitive to amplitude and phase noise. This implies that the optimal displacements, shown in Table \ref{Results:Table:amplitudes} and Fig. \ref{results:fig:settings}, are not so strict, and that slight deviations from these displacements are acceptable.\\

\section{Conclusion}
We have proposed an experiment for demonstrating nonlocality with multiple parties separated by a set of lossy channels. The experiment utilizes only standard quantum optical elements, including on/off detectors, beamsplitters, two-mode squeezers and displacements. We have given a detailed account of how loss impact the experiment, and identified critical values for channel transmissions, required for a Bell inequality violation with dichotomic inputs and outputs.
We found that the experiment is very robust against loss in the channels connecting the parties (ch$_\texttt{S}$), allowing for transmissions as low as 10\%. On the other hand, our calculations indicate that the nonlocality of the experiment is strongly impacted by loss in the channels connecting the two-mode squeezer of each party, to the detector associated with that party (channels ch$_\texttt{P}$). However, we found that the experiment could be made more robust against loss in channels ch$_\texttt{P}$, if the number of parties is increased. With 4 parties we found that the W$^3$ZB inequality could be violated for transmissions of channels ch$_\texttt{P}$ as low as 82\%. For an experiment with 4 or fewer parties, we found that the marginal outcome probabilities for all possible subgroups were inside the Bell polytope, with the used measurement settings. \\

Due to the heralded nature of the experiment, it is very sensitive toward dark-counts at the heralding detector. Our calculations indicate that the probability of a dark-count during a measurement must not be much higher than 1 in 10000, or the experiment fails. We then examined the influence of amplitude and phase noise, and found that the experiment is quite robust against these noise sources. The phase noise could be as high as several hundred milliradians, and the relative amplitude noise could be in excess of 25\%. \\

\section{Acknowledgment}
We acknowledge the support of the Danish National Research Foundation through the Center for Macroscopic Quantum States (bigQ, DNRF0142) and research grant (40864) from VILLUM
FONDEN.

%% file: appendix.tex
\section*{A1}
The state $\rho$ is generated by N two-mode squeezers, and occupy the modes $S$ and $P$. The characteristic function of $\rho$ is given by $\chi_\rho(\Lambda) = \exp[-(1/2) \Lambda^T \Omega \sigma \Omega^T \Lambda]$ where $\Lambda$ is a vector of conjugate quadratures for the modes in $S$ and $P$. We introduce the following decomposition of the covariance matrix of $\rho$,
\begin{align}
	\sigma = 
		\begin{pmatrix}
			\sigma_{P} & K_{\bar{S}} & K_{s_N} \\
			K_{\bar{S}}^T & \sigma_{\bar{S}} & C \\
			K_{s_N}^T & C^T & \sigma_{s_N}
		\end{pmatrix} .
\end{align}
We also introduce the matrices,
\begin{align}
		 K_{S} = \begin{pmatrix} K_{\bar{S}} \hspace{0.1cm}  K_{s_N} \end{pmatrix}, \ \ \sigma_S =  \begin{pmatrix} \sigma_{\bar{S}} & C \\ C^T & \sigma_{s_N} \end{pmatrix} .
\end{align}
The subscript refer to the modes described by the relevant submatrix, i.e. $\sigma_{\bar{S}}$ describes the marginal distribution of the modes $\bar{S} = S \backslash \{s_N \}$.\\
The modes in $S$ are mixed in the interferometer B, described by the Bogoliubov transformation in Eq. \ref{model:eq:bogo}. We then condition the state on obtaining a click at detector $s_N$ and no click at the remaining detectors in $S$ (this is referred to as a swap). 
If the detectors in $S$ are triggered by a dark-count with probability $p_d$, then the swap might herald success under three different conditions,
\begin{enumerate}
	\item No dark-counts occur. Light reaches detector $s_N$ and no light reaches the remaining detectors in $S$. This event is associated with the projector $\hat{\Pi}_1 = \left( \prod_{s \in \bar{S}}\ket{0}_s \hspace{-0.09cm} \bra{0} \right) (I_{s_N}-\ket{0}_{s_N} \hspace{-0.09cm} \bra{0})$.
	\item A dark-count occurs at detector $s_N$. Light reaches detector $s_N$ and no light reaches the remaining detectors in $S$. This event is associated with the projector $\hat{\Pi}_1 = \left( \prod_{s \in \bar{S}}\ket{0}_s \hspace{-0.09cm} \bra{0} \right) (I_{s_N}-\ket{0}_{s_N} \hspace{-0.09cm} \bra{0})$.
	\item A dark-count occurs at detector $s_N$. No light reaches any detectors in $S$. This event is associated with the projector $\hat{\Pi}_2 = \prod_{s \in S}\ket{0}_s \hspace{-0.09cm} \bra{0}$.
\end{enumerate}
Let $P(\hat{\Pi}_n|C)$ be understood as the probability that the event $\hat{\Pi}_n$ occur, given that detectors $S$ herald a successful swap $C$. $P(\hat{\Pi}_n) = \operatorname{Tr}\left[\hat{\Pi}_n \rho \right]$ is the prior probability that the event $\hat{\Pi}_n$ occurs. The swap then transform the state $\rho$ into the conditional state $\rho_c$ as,
\begin{align}
	&\rho \rightarrow \rho_c \nonumber \\ &= \operatorname{Tr}_{S} \left[ P(\hat{\Pi}_1|C) \frac{\hat{\Pi}_1 \rho \hat{\Pi}_1}{P(\hat{\Pi}_1)} + P(\hat{\Pi}_2|C) \frac{\hat{\Pi}_2 \rho \hat{\Pi}_2}{P(\hat{\Pi}_2)} \right] \nonumber \\
	&= \operatorname{Tr}_{S} \left[ \rho \left(  P(\hat{\Pi}_1|C) \frac{\hat{\Pi}_1}{P(\hat{\Pi}_1)} + P(\hat{\Pi}_2|C) \frac{\hat{\Pi}_2}{P(\hat{\Pi}_2)}  \right) \right] .
	\label{appendix:eq:rhoTrans1}
\end{align}
By Bayes' theorem we have,
%Examining theand introduced the operator $\widetilde{\Pi}_c$,
%\begin{align}
%	\widetilde{\Pi}_c =  P(\hat{\Pi}_1|C) \frac{\hat{\Pi}_1}{P(\hat{\Pi}_1)} + P(\hat{\Pi}_2|C) \frac{\hat{\Pi}_2}{P(\hat{\Pi}_2)} .
%\end{align}
\begin{align}
	\frac{P(\hat{\Pi}_n|C)}{P(\hat{\Pi}_n)} = \frac{P(C|\hat{\Pi}_n)}{P(C)} , 
\end{align}
which gives another expression for $\rho_c$,
\begin{align}
\rho_c &= \operatorname{Tr}_{S} \left[ \rho \left(  \frac{P(C|\hat{\Pi}_1)}{P(C)} \hat{\Pi}_1 + \frac{P(C|\hat{\Pi}_2)}{P(C)} \hat{\Pi}_2  \right) \right] \nonumber \\
&= \frac{1}{P(C)} \operatorname{Tr}_{S} \left[ \rho \widetilde{\Pi}_c \right]
\end{align}
Where we have introduced the operator $\widetilde{\Pi}_c$,
\begin{align}
	\widetilde{\Pi}_c =  P(C|\hat{\Pi}_1) \hat{\Pi}_1 + P(C|\hat{\Pi}_2) \hat{\Pi}_2
\end{align}
The probability of the swap being heralded as successful, given that the event $\hat{\Pi}_1$ occur, is given by $P(C|\hat{\Pi}_1) = (1-p_d)^N + (1-p_d)^{N-1}p_d$, i.e. the swap will succeed as long as no dark-count triggers any detector other than $s_N$. If no light reaches any detectors in $S$, then the swap can only be heralded as successful if a dark-count triggers detector $s_N$, so $P(C|\hat{\Pi}_2) = (1-p_d)^{N-1}p_d$. Then we have,
\begin{align}
	&\widetilde{\Pi}_c =  (1-p_d)^N \hat{\Pi}_1 + (1-p_d)^{N-1}p_d( \hat{\Pi}_1 + \hat{\Pi}_2 ) \nonumber \\ &= (1-p_d)^{N-1} \left( \prod_{s \in \bar{S}}\ket{0}_s \hspace{-0.09cm} \bra{0} \right) \left[  I_{s_N}-(1-p_d)\ket{0}_{s_N} \hspace{-0.09cm} \bra{0} \right]
\end{align}
Different number of photons could \textit{in principle} be distinguishable by the detector, even if the experimenter cannot distinguish the detector states sufficiently well to obtain this information. We define a projector onto Fock states, $\hat{\Pi}^{(n)}= \left( \prod_{s \in \bar{S}}\ket{0}_s \hspace{-0.09cm} \bra{0} \right) \ket{n}_{s_N} \hspace{-0.09cm} \bra{n}$. If different Fock states are in principle distinguishable, then the transformation of $\rho$, conditioned on the swap, ought to be,
\begin{align}
	\rho &\rightarrow \operatorname{Tr}_{S} \left[ \sum_{n=0}^\infty P\left(\hat{\Pi}^{(n)}|C\right) \frac{\hat{\Pi}^{(n)} \rho \hat{\Pi}^{(n)}}{P \left( \hat{\Pi}^{(n)} \right)} \right] \nonumber \\ &= \operatorname{Tr}_{S}  \left[ \rho \sum_{n=0}^\infty  \frac{P\left(\hat{\Pi}^{(n)}|C\right)}{P \left( \hat{\Pi}^{(n)} \right)} \hat{\Pi}^{(n)} \right]  
	\label{appendix:eq:rhoTrans2}
\end{align}
Using Bayes' theorem we have,
\begin{align}
	 &= \frac{1}{P(C)} \operatorname{Tr}_{S} \left[ \rho \sum_{n=0}^\infty P\left(C|\hat{\Pi}^{(n)} \right) \hat{\Pi}^{(n)} \right] \nonumber \\ 
	 &= \frac{1}{P(C)} \operatorname{Tr}_{S}  \left[ \rho \widetilde{\Pi}_c' \right]
\end{align}
We then make the assumption that,
\begin{align}
	P\left(C|\hat{\Pi}^{(n)} \right)=
	\begin{cases}
			(1-p_d)^N + (1-p_d)^{N-1}p_d, & \text{if $n>0$}\\
            (1-p_d)^{N-1}p_d, & \text{if $n=0$}
	\end{cases}
\end{align}
Under this assumption one can show that $\widetilde{\Pi}_c' = \widetilde{\Pi}_c$, and it doesn't matter whether we use the transformation in Eq. \ref{appendix:eq:rhoTrans1} or in Eq. \ref{appendix:eq:rhoTrans2}. \\
The characteristic function of $\widetilde{\Pi}_c$ is given by,
\begin{align}
	&\chi_c(\Lambda_{S}) = \operatorname{Tr}_S \left[ \widetilde{\Pi}_c D_{S}(\Lambda_{S}) \right] \nonumber \\ &= (1-p_d)^{N-1} E(\Lambda_{\bar{S}}) \cdot \left( \pi \delta^{(2)}(\Lambda_{s_N}) - (1-p_d) E(\Lambda_{s_N}) \right)
\end{align}
Then we have that,
\begin{align}
	&\rho_c = \frac{1}{P(C)} \operatorname{Tr}_{S} [\rho \widetilde{\Pi}_c] = \nonumber \\ & \frac{1}{P(C)} \int_{\mathbb{R}^{4N}} D_{P}(-\Lambda_{P}) \chi_\rho(\Lambda_{P}, \Lambda_{S}) \chi_c(-\Lambda_{S}) \frac{d^{4N}\Lambda}{\pi^{2N}} .
	\label{appendix:a1:condRho}
\end{align}
In evaluating the above expression we have used Glauber’s formula \cite{Ferraro:2005} to express $\rho$ and $\widetilde{\Pi}_c$ in terms of their characteristic functions ($\chi_\rho$ and $\chi_c$),
\begin{align}
	\hat{O} = \int_{\mathbb{R}^{2n}} \frac{d^{2n}B}{\pi^n} \chi_O(B) D^{\dagger}(B) ,
	\label{appendix:eq:glauber}
\end{align}
where $n$ is the number of modes. We also used the facts,
\begin{align}
	\operatorname{Tr}_i [D(\Lambda_i)] = \pi \delta^{(2)}(\Lambda_i)
\end{align}
\begin{align}
D(\Lambda_i) D(\Lambda_j) = D(\Lambda_i + \Lambda_j) \exp[-i\Lambda_i^T \omega \Lambda_j]
\end{align}
From Eq. \ref{appendix:a1:condRho} we may read off the characteristic function of the conditional state $\rho_c$,
\begin{align}
	\chi_{\rho_c} (\Lambda_{P}) = \frac{1}{\pi^N P(C)} \int_{\mathbb{R}^{2N}} \chi_{\rho}(\Lambda_{P}, \Lambda_{S}) \chi_c(-\Lambda_{S}) d^{2N} \Lambda_{S} .
\end{align}
Inserting the expressions for $\chi_\rho$ and $\chi_c$, we may evaluate the conditional state as,
\begin{align}	
\chi_{\rho_c}(\Lambda_{P}) = \frac{(1-p_d)^{N-1}}{P(C)} \left[ \chi_{\bar{S}}(\Lambda_{P}) - (1-p_d) \chi_{S}(\Lambda_{P}) \right] .
\end{align}
$\chi_{\bar{S}}$ and $\chi_{S}$ are Gaussian and respectively given by
\begin{align}
\chi_{\bar{S}}(\Lambda_{P}) &= \frac{1}{\pi^N} \int_{\mathbb{R}^{2N}} \chi_{\rho}(\Lambda_{P}, \Lambda_{S}) E(\Lambda_{\bar{S}}) \pi \delta^{(2)}(\Lambda_{s_N}) d^{2N} \Lambda_{S} \nonumber \\ &= 2^{N-1} ||\gamma_{\bar{S}}||^{-1/2} E\left[V_{\bar{S}},0 \right] (\Lambda_{P}) \\
	\chi_{S}(\Lambda_{P}) &= \frac{1}{\pi^N} \int_{\mathbb{R}^{2N}} \chi_{\rho}(\Lambda_{P}, \Lambda_{S}) E(\Lambda_{S})  d^{2N} \Lambda_{S}  \nonumber \\ &= 2^N ||\gamma_{S}||^{-1/2} E\left[ V_{S},0 \right](\Lambda_{P}) .
\end{align}
Where the brackets $|| . ||$ refer to the determinant and,
\begin{align}
	E \left[V, \bar{x} \right](B) &= \exp\left[ -\frac{1}{2} B^T \Omega V \Omega^T B - i (\Omega \bar{x})^T B \right], \nonumber \\
	\gamma_{\bar{S}}&=\sigma_{\bar{S}}+I, \ \ \gamma_{S}=\sigma_{S}+I, \nonumber \\
	V_{\bar{S}}&=\sigma_{P}-K_{\bar{S}}\hspace{0.09cm} \gamma_{\bar{S}}^{-1} \hspace{0.09cm} K_{\bar{S}}^T, \nonumber \\
	V_{S}&=\sigma_{P}-K_{S} \hspace{0.09cm} \gamma_{S}^{-1} \hspace{0.09cm} K_{S}^T .
\end{align}
The normalization $P(C)$ can be obtained by demanding that $\chi_{\rho_c}(\Lambda_{P} = 0) = 1$. $E \left[V, \bar{x} \right](B)$ is the characteristic function of a Gaussian state with covariance matrix $V$ and centred on position $\bar{x}$ in phase space.\\
 
We now derive a closed-form expression for the correlator $\langle \prod_{p \in P} M_p^{(n_p)} \rangle$, describing correlations between the measurement outcomes obtained by the N parties. The characteristic function of the observable $\prod_{p \in P} M_p^{(n_p)}$ is given by,
\begin{align}
& \chi_M\left(\Lambda, X_P\right) \nonumber \\ 
& =\prod_{p \in P}\left\{\pi \delta^{(2)} \left( \Lambda_p \right)-2 (1-p_d) E\left[I, -2X_p^{(n_p)}\right]\left(\Lambda_p\right) \right\} .
\end{align}
%\begin{align}
%& \chi_M\left(\Lambda, X_P\right) \nonumber \\ 
%&=\operatorname{Tr}\left\{\hat{\Pi}_p^{2 N}\left(I_p-2(1-p_d)\left|-X_p^{(n_p)%}\right\rangle_p\left\langle-X_p^{(n_p)} \right| \right) D_p%\left(\Lambda_p\right)\right\}
%\nonumber \\
%& =\hat{\Pi}_p^{2 N}\left\{\pi \delta^{(2)} \left( \Lambda_p \right)-2 (1-p_d) %E\left[I, -X_p^{(n_p)}\right]\left(\Lambda_p\right) \right\} .
%\end{align}
As we will show in the next section, when amplitude or phase noise is present, then we should instead use the characteristic function,
\begin{align}
	&\chi_M\left(\Lambda_P, X_P\right)  \nonumber \\
	&=\prod_{p \in P}\left\{\pi \delta^{(2)} \left( \Lambda_p \right)-2 (1-p_d) E\left[\Delta_p^{(n_p)}, -2X_p^{(n_p)}\right]\left(\Lambda_p\right) \right\},
\end{align}
where $\Delta_p^{(n_p)}$ is the covariance matrix describing a noisy displacement for party $p$. We form the covariance matrix $\Delta_{P}$, describing the statistics of the noisy displacements for all N modes. We assume no correlation between noise in different modes, and $\Delta_{P}$ is therefore block diagonal. The above product is rewritten as a sum over products,
\begin{align}
\chi_M\left(\Lambda_P, X_P\right)=\sum_d [-2(1-p_d)]^{|d|} \prod_{p \in P} K_p^{\left(d_p\right)} ,
\end{align}
where the sum runs over all binary lists $d=\left(d_{p_1}, d_{p_2}, \ldots, d_{p_N}\right)$. $|d|$ is the sum of $d$, i.e. the number of ones in the list. $K_p^{\left(d_p\right)}$ is the piecewise characteristic function defined as,
\begin{align}
K_p^{\left(d_p\right)}= \begin{cases}\pi \delta^2\left(\Lambda_p\right) & \text { if } d_p=0 \\ E\left[\Delta_p^{(n_p)}, -2X_p^{(n_p)}\right]\left(\Lambda_p\right) & \text { if } d_p=1\end{cases} .
\end{align}
Given a Gaussian state $\rho_G$ with characteristic function $E[\sigma_G, 0](\Lambda_P)$, we evaluate the expectation value of the observable,
\begin{align}
&f\left(\sigma_G, X_P\right)=\operatorname{Tr}\left\{\rho_G \prod_{p \in P} M_p^{\left(n_p\right)}\right\}\nonumber \\ &=\frac{1}{\pi^{N}} \int_{\mathbb{R}^{2N}} E[\sigma_G, 0](-\Lambda_P) \chi_M\left(\Lambda_P, X_{P}\right) d^{2 N} \Lambda_P \nonumber \\
&=\sum_d [-8 \pi (1-p_d)]^{|d|} G\left[\sigma_G^{(d)}+\Delta_P^{(d)},0\right]\left(2 X_P^{(d)}\right)
\end{align}
$\sigma_G^{(d)}$ is the submatrix of $\sigma_G$ containing all the modes where $d$ is 1, i.e. if $d=(1,0,1,1)$ then we extract the covariance matrix describing the marginal distribution of modes $p_1$, $p_3$ and $p_4$. Likewise, we have for the present example $\Delta_P^{(d)} = \operatorname{Diag}\left(\Delta_{p_1}^{(n_{p_1})}, \Delta_{p_3}^{(n_{p_3})}, \Delta_{p_4}^{(n_{p_4})} \right)$ and $X_P^{(d)}=X_{p_1}^{(n_{p_1})} \bigoplus X_{p_3}^{(n_{p_3})} \bigoplus X_{p_4}^{(n_{p_4})} $. We have also defined the normal distribution, $G[V,\bar{x}](X)=\left[(2 \pi)^D\|V\|\right]^{-1 / 2} e^{-\frac{1}{2} \left(X-\bar{x}\right)^T V^{-1} \left(X-\bar{x}\right)}$, where $D$ is the dimension of $V$.
Applying this result to the conditional state, which is a sum of two Gaussians, we obtain
\begin{align}
\left\langle\prod_{p \in P} M_p^{\left(n_p\right)}\right\rangle &=\operatorname{Tr}\left\{\rho_c \prod_{p \in P} M_p^{\left(n_p\right)}\right\} \nonumber \\
&=\frac{(1-p_d)^{N-1}}{P(C)} \left[ 2^{N-1}\left\|\gamma_{\bar{S}}\right\|^{-\frac{1}{2}} f\left(V_{\bar{S}}, X_P\right) \right. \nonumber \\  &\left. - 2^{N} (1-p_d)\left\|\gamma_S\right\|^{-\frac{1}{2}} f\left(V_S, X_P \right) \right] .
\end{align}
Which is a closed-form expression for the correlator of the measurements.

\subsection*{Loss}
A Gaussian transformation transforms the quadrature operators as $Q \rightarrow SQ + d$, where $S$ is a symplectic matrix, i.e. $S \Omega S^T = \Omega$, and $d$ is a displacement \cite{Ferraro:2005,Weedbrook:2012}. Correspondingly, one can show that under a Gaussian transformation, the characteristic function transforms as,
\begin{align}
	\chi(\Lambda) \rightarrow \exp\left[ i d^T \Omega \Lambda \right] \chi(S^{-1} \Lambda) .
	\label{appendix:eq:charTrans}
\end{align}
We note that $S^{-1} = \Omega^T S^T \Omega$. We model loss, acting on the optical modes of the system, by mixing said modes with a set of empty (groundstate) environmental modes, and subsequently trace out the environmental modes. Let the modes be ordered as $\Lambda = \Lambda_P \oplus \Lambda_S \oplus \Lambda_E$, where $\Lambda_E$ are the conjugate quadratures for the environmental modes. We assume there is one environmental mode for each system mode ($S$, $P$). The system modes and environmental modes are mixed using beamsplitter interactions, described by the symplectic matrix $U_\eta$,
\begin{align}
	U_\eta = 
		\begin{pmatrix}
			G^{1/2}_\eta & - \sqrt{I-G_\eta}\\
			\sqrt{I-G_\eta} & G^{1/2}_\eta
		\end{pmatrix},
\end{align}
By using Eq. \ref{appendix:eq:glauber}, Eq. \ref{appendix:eq:charTrans}, and $U_\eta$, we obtain the map corresponding to loss acting on the system modes. This map transforms the characteristic function as,
\begin{align}
	\chi(\Lambda) \rightarrow \chi \left( G^{1/2}_\eta \Lambda \right) \exp \left[ -\frac{1}{2} \Lambda^T (I-G_\eta) \Lambda \right] ,
\end{align}
Eq. \ref{model:eq:lossMap} can be derived from this mapping, and it can also be used to show that detector loss can be commuted through the interferometer B, given that all detectors have the same efficiency.

\subsection*{Phase and amplitude noise}
We now evaluate the effect of phase and amplitude noise on the computed correlators. Given that the optical state $\rho$ is perturbed in phase by the environment, we model this by stochastic rotations in phase space $
\rho=\int d^{N} \boldsymbol{\theta} P(\boldsymbol{\theta}) R(\boldsymbol{\theta}) \rho_0 R(-\boldsymbol{\theta})$. Where $\rho_0$ is the unperturbed state, $\boldsymbol{\theta}=\begin{pmatrix}\theta_{p_1} \theta_{p_2} \ldots \theta_{p_N} \end{pmatrix}$ is a vector of stochastic rotation angles, and $R(\boldsymbol{\theta})$ is the rotation operator $R(\boldsymbol{\theta})=\prod_{p \in P} R_p\left(\theta_p\right)$. We shift this stochastic rotation from the state onto the observable:
\begin{align}
&\left\langle\prod_{p \in P} M_p^{\left(n_p\right)}\right\rangle =  \operatorname{Tr}\left\{\prod_{p \in P} M_p^{\left(n_p\right)} \rho\right\} \nonumber \\ &=\operatorname{Tr}\left\{\prod_{p \in P} M_p^{\left(n_p\right)} \int d^{N} \boldsymbol{\theta} P(\boldsymbol{\theta}) R(\boldsymbol{\theta}) \rho_0 R(-\boldsymbol{\theta})\right\} \nonumber \\
&=\operatorname{Tr}\left\{\int d^{N} \boldsymbol{\theta} P(\boldsymbol{\theta}) R(-\boldsymbol{\theta}) \prod_{p \in P} M_p^{\left(n_p\right)} R(\boldsymbol{\theta}) \rho_0\right\} \nonumber \\
&= \operatorname{Tr}\left\{\prod_{p \in P} \int d \theta_p P\left(\theta_p\right) R_p\left(-\theta_p\right) M_p^{\left(n_p\right)} R_p\left(\theta_p\right) \rho_0\right\} \nonumber \\
&= \operatorname{Tr}\left\{\prod_{p \in P} \widetilde{M}_p^{\left(n_p\right)} \rho_0\right\}
\end{align}
Where $\widetilde{M}_p^{\left(n_p\right)}$ is the noisy observable. By factorizing the probability as $P(\boldsymbol{\theta})=\prod_{p \in P} P(\theta_p)$, we have tacitly assumed that there is no correlation in the phase noise acting on different modes. Inserting the expression for the observable $M_p^{\left(n_p\right)}$, we have
\begin{align}
	&R_p\left(-\theta_p\right) M_p^{\left(n_p\right)} R_p\left(\theta_p\right) \nonumber \\ &=  I_p - 2 \left(1-p_d\right) R_p\left(-\theta_p\right)\left|-X_p^{\left(n_p\right)}  \right\rangle_p \hspace{-0.09cm} \left\langle-X_p^{\left(n_p\right)}\right| R_p\left(\theta_p\right)
\end{align}
For a coherent state $\ket{-X_p^{(n_p)}}$, we have that a small rotation is identical to a displacement acting orthogonal to the amplitude vector $-X_p^{\left(n_p\right)}$. An orthogonal vector can be constructed by acting with the symplectic form: $-\omega (-X_p^{\left(n_p\right)})$. With this in mind, we make the substitution:
\begin{align}
R_p\left(\theta_p\right) \rightarrow D_p\left(\theta_p \omega X_p^{(n_p)}\right)
\end{align}
Imprecision in the measurement process, such as a noisy displacement, might lead to noise in the amplitude. We include this by also applying a stochastic displacement along the amplitude vector $X_p^{\left(n_p\right)}$. This stochastic displacement is given as a fraction $r_p$ of the amplitude vector $X_p^{\left(n_p\right)}$, i.e. the stochastic displacement is $r_p X_p^{\left(n_p\right)}$. $r_p$ is referred to as the relative amplitude. The noisy observable for party $p$ is then given as,
\begin{widetext}
\begin{align}
\widetilde{M}_p^{\left(n_p\right)} & =\int d \theta_p d r_p P\left(\theta_p, r_p\right) D_p\left(-\theta_p \omega X_p^{(n_p)}\right) D_p\left(-r_p X_p^{\left(n_p\right)}\right) M_p^{\left(n_p\right)} D_p\left(r_p X_p^{\left(n_p\right)}\right) D_p\left(\theta_p \omega X_p^{(n_p)}\right) \nonumber \\
&=I-2 \left(1-p_d\right) \int P\left(\theta_p, r_p\right)\cdot \left| -\left(1+r_p + \theta_p \omega \right) X_p^{(n_p)} \right\rangle \left\langle -\left(1+r_p + \theta_p \omega \right) X_p^{(n_p)} \right| d \theta_p d r_p \nonumber \\
&= I-2 (1-p_d)\beta_p^{(n_p)}
\end{align}
$P\left(\theta_p, r_p\right)$ is the distribution over displacements, and we have introduced the state,
\begin{align}
\beta_p^{\left(n_p\right)}=\int P\left(\theta_p, r_p\right) \cdot\left| -\left(1+r_p + \theta_p \omega \right) X_p^{(n_p)} \right\rangle \left\langle -\left(1+r_p + \theta_p \omega \right) X_p^{(n_p)} \right| d \theta_p d r_p .
\end{align}
We model $P\left(\theta_p, r_p\right)$ as a Gaussian, given by
\begin{align}
&P\left(\theta_p, r_p\right) 
=\left[(2 \pi)^2\left\|\Sigma_p\right\|\right]^{-1/2} \exp\left[-\frac{1}{2} \begin{pmatrix} r_p & \theta_p \end{pmatrix} \Sigma_p^{-1} \begin{pmatrix} r_p \\ \theta_p \end{pmatrix} \right] .
\end{align}
The covariance matrix is chosen to be diagonal
\begin{align}
\Sigma_p=\left(\begin{array}{cc}
V_A & 0 \\
0 & V_\theta
\end{array}\right) .
\end{align}
$V_A$ and $V_\theta$ are the relative amplitude and phase angle variance respectively. $\beta_p^{\left(n_p\right)}$ has a characteristic function given by,

\begin{align}
\chi_{\beta_p^{\left(n_p\right)}} &= \operatorname{Tr}\left\{\beta_p^{\left(n_p\right)} D_p\left(\Lambda_p\right)\right\} \nonumber \\
&=\int P\left(\theta_p, r_p\right) \cdot E\left[I,-\left(1+r_p + \theta_p \omega \right) 2 X_p^{(n_p)}\right]\left(\Lambda_p\right) d \theta_p d r_p \nonumber \\
&=E\left[I+V_A \left( 2 X_p^{\left(n_p\right)} \right) \otimes \left(2 X_p^{(n_p)}\right)^T+V_\theta \left( \omega^T 2 X_p^{(n_p)} \right) \otimes \left(\omega^T 2 X_p^{(n_p)}\right)^T ,-2 X_p^{\left(n_p\right)}\right]\left(\Lambda_p\right) .
\end{align}
\end{widetext}
So the effect of amplitude and phase noise is to broaden the phase space distribution of $\beta_p^{\left(n_p\right)}$ along $2X_p^{(n_p)}$ and $\omega^T 2 X_p^{(n_p)}$. We define the covariance matrix of the state $\beta_p^{\left(n_p\right)}$ as $\Delta_p^{\left(n_p\right)}$,
\begin{align}
\Delta_p^{\left(n_p\right)}&=I+V_A \left( 2 X_p^{\left(n_p\right)} \right) \otimes \left( 2 X_p^{(n_p)}\right)^T \nonumber \\ 
&+V_\theta \left( \omega^T 2 X_p^{(n_p)} \right) \otimes \left(\omega^T 2 X_p^{(n_p)}\right)^T
\end{align}

\section*{A2}
Let $n$ be a binary list of measurement settings, and $g$ a binary list of measurement outcomes for the detectors in $P$, where click corresponds to 1 and no click corresponds to 0. We may then compute the probability of obtaining the outcomes $g$ using the characteristic function $\chi_{\rho_c}$. This probability is given by the expression,
\begin{align}
P_Q(g|n) &=\frac{(1-p_d)^{N-1}}{P(C)}\left[2^{N-1}\left\|\gamma_{\bar{S}}\right\|^{-\frac{1}{2}} h_g\left(V_{\bar{S}}\right) \right. \nonumber \\ &\left. - 2^{N}\left(1-p_d\right)\left\| \gamma_S \right\|^{-\frac{1}{2}} h_g\left(V_S\right)\right],
\label{a2:eq:pq}
\end{align}
where
\begin{align}
h_g(V)&=\left[4 \pi\left(1-p_d\right)\right]^{|\bar{g}|} \sum_b\left[-4 \pi\left(1-p_d\right)\right]^{|b|} G\left[V^{(b+\bar{g})} \right. \nonumber \\  &\left. +\Delta_P^{(b+\bar{g})}, 2 X_P^{(b+\bar{g})}\right] .
\end{align}
$\bar{g}$ is the negation of $g$, i.e. we replace 1 by 0 and vice versa. The measurement settings $n$ define the arrays $\Delta_{P}$ and $X_P$. The sum runs over all binary lists $b$ of length N, satisfying the constraint that $b$ takes the value zero in positions where $g$ takes the value zero. E.g. if $g = \begin{pmatrix} 1, 0, 0, 1 \end{pmatrix}$, then the sum would run over the lists $b \in \left\{ \begin{pmatrix} 0, 0, 0, 0 \end{pmatrix},\begin{pmatrix} 1, 0, 0, 0 \end{pmatrix},\begin{pmatrix} 0, 0, 0, 1 \end{pmatrix},\begin{pmatrix} 1, 0, 0, 1 \end{pmatrix} \right\}$. $V^{(b+\bar{g})}$ is the submatrix of the covariance matrix $V$, containing all the modes where the vector $b+\bar{g}$ takes the value 1, e.g. if $b+\bar{g}=\begin{pmatrix} 0, 1, 1, 1 \end{pmatrix}$ then the marginal covariance matrix describing modes $p_2$, $p_3$ and $p_4$ is extracted. Marginal probabilities for a subset of parties A can be extracted from $P_Q(g|n)$ by summing over outcomes for the remaining parties B. The measurement settings for subset B should be fixed during this summation, however the choice of settings for B is arbitrary owing to the no-signalling property of quantum mechanics \cite{Brunner:2014}.\\

We then want to determine whether the array $P_Q(g|n)$ can be expressed as a convex sum of local response functions. Let $L\left (g_p|n_p,\lambda_k \right)$ be the local response function for party $p$, determined by the hidden variables $\lambda_k$. The response function gives the probability of party $p$ obtaining a particular outcome $g_p$, given the measurement setting $n_p$ and hidden variables $\lambda_k$. We determine whether there exists a set of coefficients $c_k$ such that \cite{Brunner:2014}:
\begin{align}
P_Q(g|n) &= \sum_{k} c_k \prod_{p \in P} L\left (g_p|n_p,\lambda_k \right) \nonumber \\
\sum_k c_k &= 1  \nonumber \\
c_k  &\geq 0
\label{a2:eq:psum}
\end{align}
$c_{k}$ is interpreted as the probability that the hidden variables $\lambda_k$ are shared by the parties in a given measurement round. We use the set of deterministic response functions, i.e. each response function can be written as a Kronecker delta function,
\begin{align}
	L\left (g_p|n_p,\lambda_k \right) = \delta({g_p,g_{n_p,\lambda_k}})
\end{align}
$g_p$ is a potential outcome for party $p$ and $g_{n_p,\lambda_k}$ is the outcome that is actually obtained, given the hidden variables $\lambda_k$ and the setting $n_p$. Whether the set of requirements in Eq. \ref{a2:eq:psum} allows for a solution or not, is determined using the linprog module of the SciPy 1.8.1 package in Python. When no solution is present, we know that the array of probabilities $P_Q(g|n)$, determined by the quantum state, does not admit a local hidden variable model. In this case $P_Q(g|n)$ lies outside the Bell polytope. However, when a solution \textit{is} present we know that the system can be described by a local hidden variable model, and no Bell inequality can be violated.
%If we divide the parties into two groups, subgroup A and subgroup B. The marginal outcome probabilities for subgroup A is obtained by summing over outcomes in subgroup B, for a fixed choice of settings in subgroup B.
%\begin{align}
%&P^{(A)}_Q(\mathbf{g}^{(A)}|\mathbf{n}^{(A)}) = P_A(\mathbf{g}^{(A)}|\mathbf{n}^{(A)}) = %\sum_{k} c_k^{(A)} \prod_{p \in P^{(A)}} L\left( g_p | n_p, \lambda \right) \nonumber \\
%&\sum_k c_k = 1  \nonumber \\
%&c_k  \geq 0
%\end{align}